\documentclass[aps,prd,twocolumn,preprintnumbers,superscriptaddress,dblfloatfix,nofootinbib]{revtex4-1}
\usepackage{graphicx}   
\usepackage{dcolumn}    
\usepackage{bm}         
\usepackage{amssymb}    
\usepackage{setspace}
\usepackage{amsmath, amssymb, setspace}
\usepackage{array}
\usepackage{booktabs}
\usepackage{mathrsfs}
\usepackage{indentfirst}
\usepackage{slashed}
\usepackage{float}
\usepackage{lmodern}
\usepackage{hyperref}
\usepackage{xcolor}
\usepackage{multirow}
\usepackage{epstopdf}
\usepackage{ulem}
\usepackage{tabularx}
\usepackage[justification=raggedright]{caption}
\usepackage[flushleft]{threeparttable}
\usepackage{hyperref}
\usepackage{soul}
\usepackage[utf8]{inputenc}

\begin{document}


\title{Visible Sterile Neutrinos as the Earliest Relic Probes of Cosmology}

\author{Graciela B. Gelmini}
\email{gelmini@physics.ucla.edu}
\affiliation{Department of Physics and Astronomy, University of California, 
Los Angeles, CA 90095-1547, USA}

\author{Philip Lu}
 \email{philiplu11@gmail.com}
\affiliation{Department of Physics and Astronomy, University of California, 
Los Angeles, CA 90095-1547, USA}
 
\author{Volodymyr Takhistov}
\email{vtakhist@physics.ucla.edu}
\affiliation{Department of Physics and Astronomy, University of California, 
Los Angeles, CA 90095-1547, USA} 

\date{\today}

\begin{abstract} 
A laboratory detection of a  sterile neutrino could provide the first indication of the evolution of the Universe before Big-Bang Nucleosynthesis (BBN), an epoch yet untested.  Such ``visible" sterile neutrinos are  observable in upcoming experiments such as KATRIN/TRISTAN and HUNTER in the keV mass range and PTOLEMY and others in the eV mass range.   A set of standard assumptions is typically made about cosmology before the temperature of the Universe was 5 MeV. However, non-standard pre-BBN cosmologies based on alternative assumptions could arise in motivated theoretical models and are equally in agreement with all existing data. We revisit the production of sterile neutrinos of mass  0.01 eV to 1 MeV in two examples of such models: scalar-tensor and low reheating temperature pre-BBN cosmologies. In both of them, the putative 3.5 keV X-ray signal line corresponds to a sterile neutrino with a mixing large enough to be tested in upcoming laboratory experiments. Additionally, the cosmological/astrophysical upper limits on active-sterile neutrino mixings are significantly weaker  than in the standard pre-BBN cosmology, which shows that these limits are not robust. For example, in the scalar-tensor case the potential signal  regions implied by the LSND and MiniBooNE short-baseline as well as  the DANSS and NEOS reactor experiments are  entirely not bound by cosmological restrictions. Our work highlights that sterile neutrinos  may constitute a sensitive probe of the pre-BBN epoch.
\end{abstract}

\maketitle

The cosmological evolution of the Universe before its temperature was $T= 5$ MeV is unknown because we have not detected so far any remnant from it.  The earliest cosmological remnants are so far the light nuclei produced during Big Bang Nucleosynthesis (BBN)  and if the highest temperature of the radiation-domination epoch in which BBN happened was just 5 MeV, BBN and all the subsequent evolution of the Universe would be unchanged~\cite{deSalas:2015glj, Hasegawa:2019jsa, DeBernardis:2008zz, Hannestad:2004px, Kawasaki:2000en, Kawasaki:1999na}. Assumptions are made about cosmology to compute the relic abundance and momentum distributions of dark matter (DM) particles which are produced in this pre-BBN epoch.
The standard assumptions are that the Universe was radiation-dominated, that only Standard Model (SM) particles are present and that no extra entropy in matter and radiation is produced. We call this set of assumptions the standard pre-BBN cosmology, which is an extension at higher temperatures of the standard cosmology we know at lower temperatures, $T < 5$ MeV.
However, cosmologies based on alternative assumptions are equally in agreement with all existing data. The pre-BBN cosmological evolution could drastically differ from standard in some well motivated theoretical models, e.g. some based on moduli, extra dimensions or quintessence. A non-standard cosmological evolution, consistent with all existing bounds, could drastically affect the properties of any relics produced before the temperature of the Universe was 5 MeV. Detecting any relics sensitive to the pre-BBN cosmological history will open a new window into this yet unexplored epoch.   
The SM predicts that  the three active neutrinos $\nu_\alpha$, $\alpha = e, \mu, \tau$, coupled to the W and Z weak gauge bosons are massless. The discovery of neutrino oscillations~\cite{Fukuda:1998mi} confirmed that neutrinos are massive and led to
detailed studies of neutrino scenarios beyond the three-flavor paradigm of the  SM.
A promising explanation of the non-zero active neutrino masses involves minimally extending the SM to include one or more additional ``sterile'' neutrinos $\nu_s$ which do not interact weakly and in minimal scenarios only mix with 
the $\nu_{\alpha}$. For simplicity, here we assume  a  $\nu_s$ that has a mixing $\sin \theta$ with only one $\nu_{\alpha}$, and we take $\alpha= e$ in our figures. Sterile neutrinos with mass of $\mathcal{O}$(keV) and a spectrum close to thermal constitute a viable Warm DM (WDM) candidate (see e.g.~\cite{Boyarsky:2018tvu}).

Sterile neutrinos without additional interactions beyond the SM are produced in the early Universe through active-sterile  flavor oscillations and collisional processes. In the absence of a large lepton asymmetry the oscillations are non-resonant, and the  analytic solution for the relic number density of sterile neutrinos  produced through this mechanism was first obtained by Dodelson and Widrow (DW)~\cite{Dodelson:1993je}. In the presence of a large lepton asymmetry sterile neutrinos could be produced via resonant conversion~\cite{Shi:1998km}. In more complicated scenarios, other production mechanisms, such as heavy scalar decays~\cite{Petraki:2007gq}, are also possible. Here  we focus on the DW production mechanism, and leave details of the results we present, as well as the study of resonant production, for future work~\cite{Gelmini:prep}.

\begin{figure*}[htb]
\begin{center}
\includegraphics[trim={5mm 0mm 40 0},clip,width=.325\textwidth]{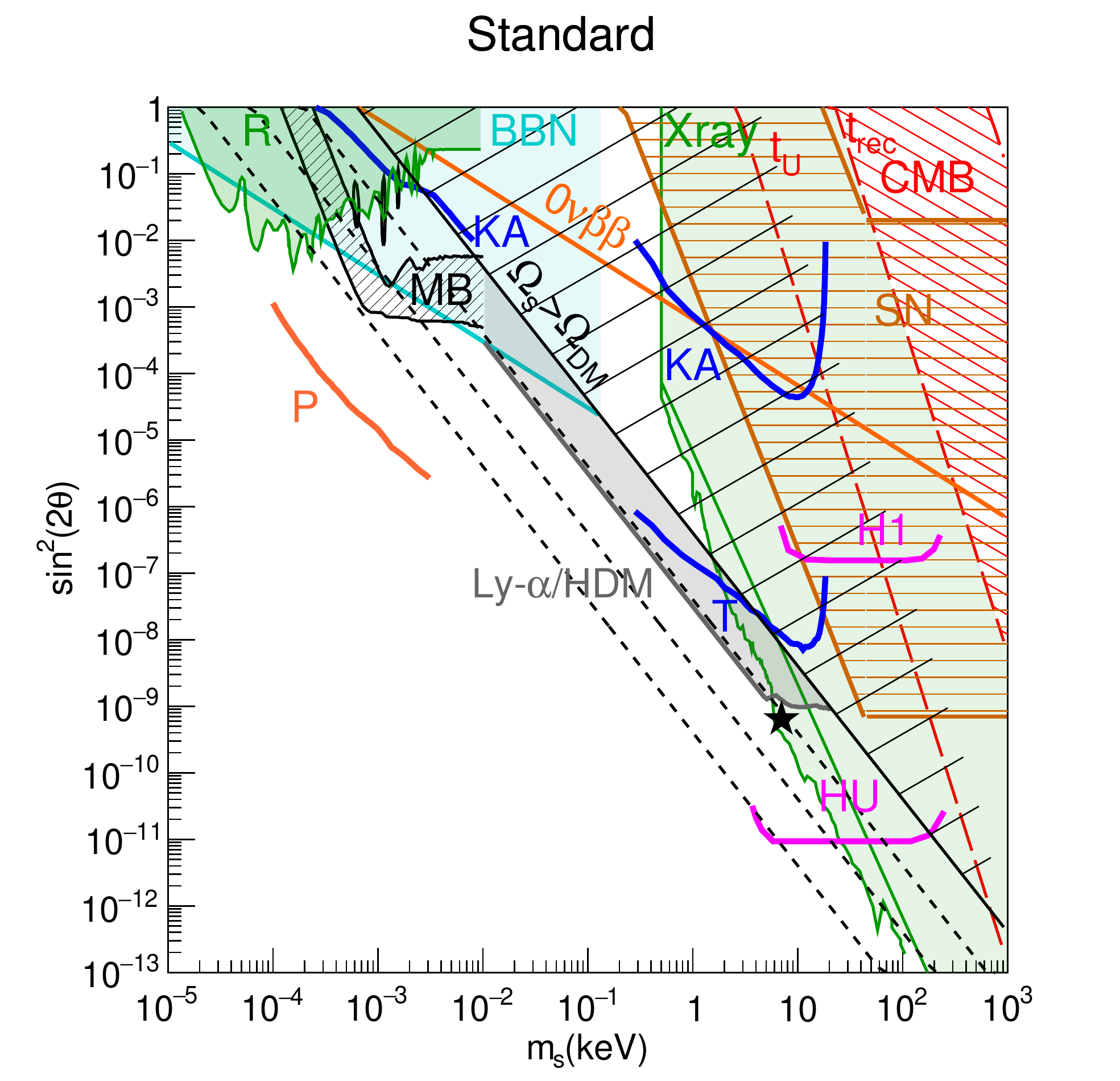}
\includegraphics[trim={5mm 0mm 40 0},clip,width=.325\textwidth]{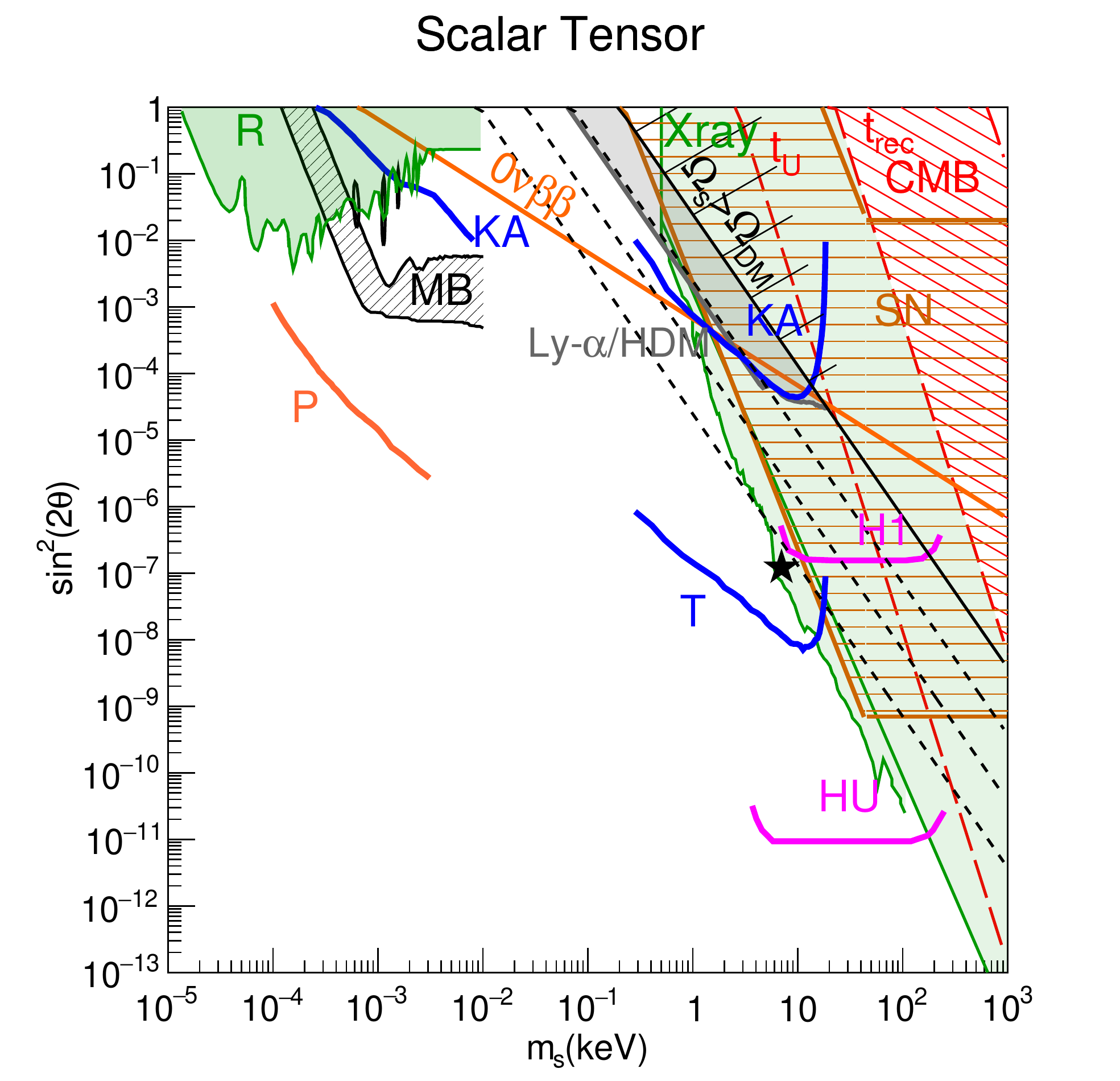}
\includegraphics[trim={5mm 0mm 40 0},clip,width=.325\textwidth]{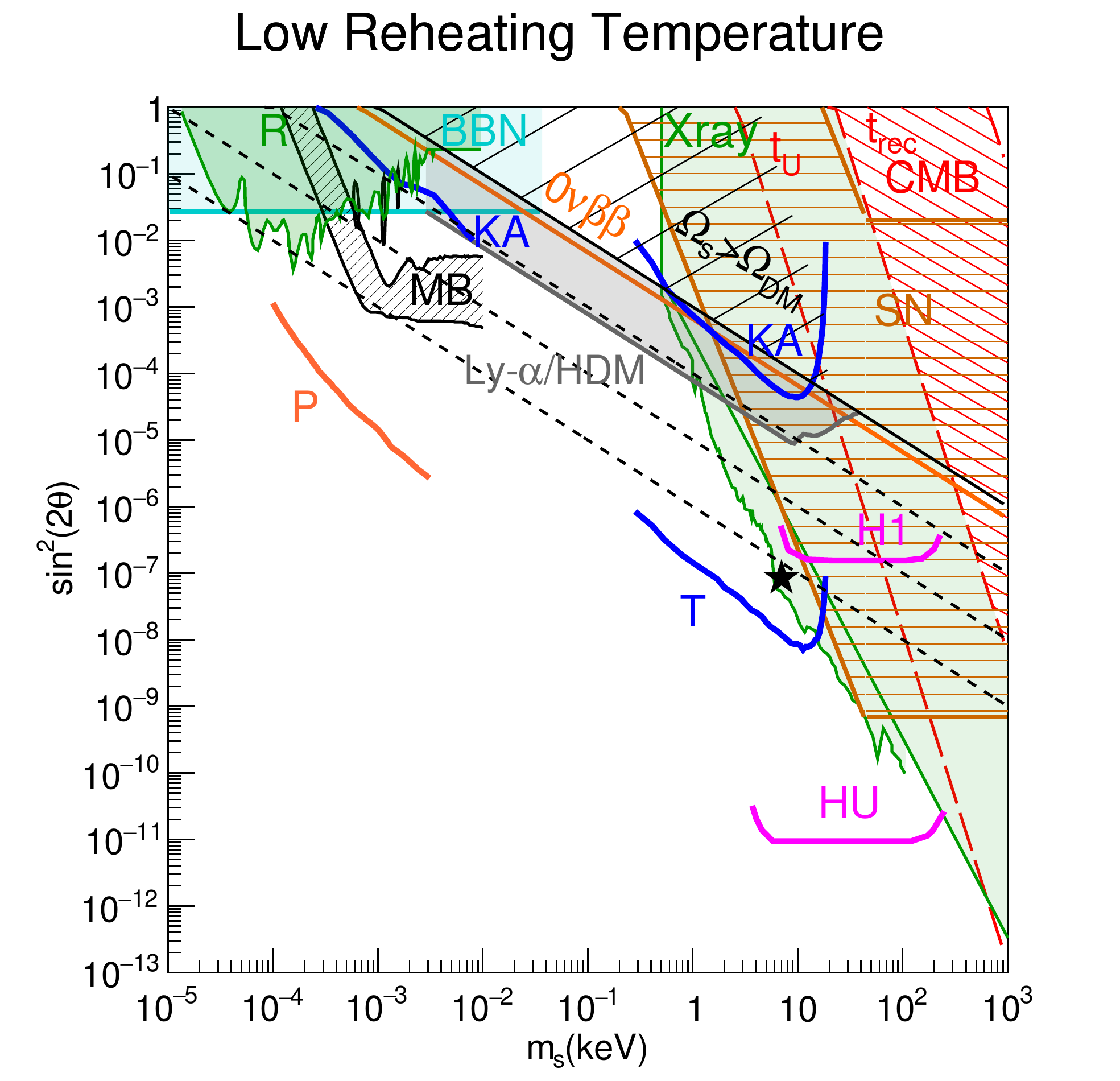}
\caption{\label{fig:allDWlim}
Present relic abundance, limits and regions of interest in the mass-mixing space of a  sterile neutrino mixed only with $\nu_e$,   assuming different pre-BBN cosmologies (indicated above each panel). The scalar-tensor and low reheating temperature cosmologies (center and right) transition to the standard cosmology at a temperature of 5 MeV.   Sterile neutrino fractional contribution to the DM of  1, and then $10^{-1}$, $10^{-2}$ and $10^{-3}$ are  shown with a black solid line and dotted lines, respectively (the forbidden region $\Omega_s/\Omega_{\rm DM} > 1$ is  diagonally hatched in black). Lifetimes $\tau= t_U$, $t_{\rm rec}$ and $t_{\rm th}$ (see text) of Majorana sterile neutrinos, are shown with straight long dashed red lines.  The region disfavored by supernovae~\cite{Kainulainen:1990bn} (SN) is horizontally hatched in brown.
 The black star shows the location  of the 3.5 keV X-ray signal~\cite{Bulbul:2014sua,Boyarsky:2014jta} for each cosmology. Regions rejected by limits from combined reactor neutrino (R) experiments (Daya Bay~\cite{An:2016luf}, Bugey-3~\cite{Declais:1994su} and  PROSPECT~\cite{Ashenfelter:2018iov}) shown in green, on $N_{\rm eff}$ during BBN~\cite{Tanabashi:2018oca} (BBN) in cyan, from
 Lyman-alpha data~\cite{Baur:2017stq} (Ly-$\alpha$/HDM) in gray, from X-ray~\cite{Ng:2019gch,Perez:2016tcq,Neronov:2016wdd} including DEBRA observations~\cite{Boyarsky:2005us}  (Xray) in green, on $0\nu\beta\beta$ decays~\cite{KamLAND-Zen:2016pfg}  ($0\nu\beta\beta$) in orange  and on CMB spectrum distortions~\cite{Fixsen:1996nj} (CMB) diagonally hatched in red. Current/future sensitivity of KATRIN (KA) in the keV~\cite{Mertens:2018vuu} and  eV~\cite{megas:thesis}  mass range and its TRISTAN upgrade in 3 years (T)~\cite{Mertens:2018vuu} are shown by blue solid lines. The magenta solid lines  show the sensitivity limits of the  phase 1B (H1) of the proposed HUNTER experiment, and its upgrade (HU)~\cite{Smith:2016vku}.   The region densely hatched in black (MB)  is the 4-$\sigma$ band of compatibility with LSND and MiniBooNE results in Fig.~4 of~\cite{Aguilar-Arevalo:2018gpe}. The three black vertical elliptical  contours above the MB band are the regions allowed at 3-$\sigma$ by DANSS~\cite{Alekseev:2018efk} and NEOS~\cite{Ko:2016owz} data in Fig.~4 of~\cite{Gariazzo:2018mwd}). An orange solid line
 shows the sensitivity projection of PTOLEMY  for
100 g-yr (P)
(see Figs. 6 and 7 of~\cite{Betti:2019ouf}).
}
\end{center}
\end{figure*}

We revisit the effects of different pre-BBN cosmologies on visible sterile neutrinos with mass $10^{-2}~\text{eV} < m_s < 1~\text{MeV}$, in view of the reported 3.5 keV X-ray line, which could be due to a 7 keV mass sterile neutrino, and  of recent results  from  short-baseline and reactor neutrino experiments, which could  be due to a sterile neutrino with $\mathcal{O}$(eV) mass. We show that, if observed
in laboratory experiments, sterile neutrinos could act as sensitive relic probes of the pre-BBN cosmology. For concreteness, we discuss sterile neutrino production within two representative non-standard pre-BBN cosmological scenarios, scalar-tensor (ST) and low-reheating temperature (LRT) cosmologies, and compare them with the standard cosmology (Std). Within the ST or LRT scenarios, the sterile neutrino production is suppressed with respect to that in the Std. As a consequence the cosmological/astrophysical upper limits on the active-sterile neutrino mixing can be significantly less restrictive than within the Std. Hence, these limits do not reject the sterile neutrino masses and mixings compatible with the reported results of the short-baseline LSND~\cite{Aguilar:2001ty} and MiniBooNE~\cite{Aguilar-Arevalo:2018gpe} (see the band labeled MB in Fig.~\ref{fig:allDWlim}) as well as the reactor neutrino DANSS~\cite{Alekseev:2018efk} and NEOS~\cite{Ko:2016owz} experiments (see three black elliptical vertical contours above the MB band in Fig.~\ref{fig:allDWlim}).
Also, the sterile neutrino responsible for the putative 3.5 keV X-ray signal line~\cite{Bulbul:2014sua,Boyarsky:2014jta} can have a much larger mixing (indicated by a black star in Fig.~\ref{fig:allDWlim}), that is within the sensitivity reach of TRISTAN~\cite{Mertens:2018vuu} and HUNTER~\cite{Smith:2016vku} (shown respectively in blue and magenta in Fig.~\ref{fig:allDWlim}).  

Several related studies have already been carried out
~\cite{Abazajian:2017tcc,Gelmini:2004ah,Rehagen:2014vna}. We extend the results of the previous LRT analysis focusing on keV-mass neutrinos of~\cite{Abazajian:2017tcc} down to 0.01 eV masses and update the older constraints of~\cite{Gelmini:2004ah}. For ST, we extend the results of~\cite{Rehagen:2014vna}, pointing out that the 3.5 keV line will be more visible and highlighting the significance of upcoming laboratory experiments.

Compared to the standard Hubble expansion rate $H_{\rm Std}$,
the ST expansion rate is $H_{\rm ST}= \eta (T/ T_{\rm tr})^\beta H_{\rm Std}$,  where $T$ is the temperature of the radiation bath, $T_{\rm tr}$ is the temperature at which there is a fast transition to the Std cosmology, and we take $\eta= 7.4 \times 10^5$ and $\beta = -0.82$ adopting the model of~\cite{Catena:2004ba} (models with different $\eta$ and $\beta$ are also possible, e.g.~\cite{Catena:2007ix}).  
In this model the entropy in matter and radiation is conserved. We require that the standard expansion rate is recovered for $T < T_{\rm tr} = 5$ MeV.
This  is  a  simplification which ensures that while our scenario follows the  behavior  of  the  scalar-tensor  model  before  BBN, it is fully compatible with all cosmological and astrophysical constraints (e.g.~\cite{Sakstein:2017xjx}).

In the LRT scenario,  entropy is not conserved in the non-standard phase~\cite{Gelmini:2004ah,Gelmini:2008fq} (see discussion in~\cite{Gelmini:2010zh}). In this phase the oscillations of a scalar field around its minimum dominate the energy density of the Universe, while the scalar field decays into radiation. Finally, the Universe becomes dominated by a radiation bath with a reheating temperature that we assume here to be $T_{RH} = 5$ MeV. This naturally can happen within models with decaying moduli fields  or low-temperature inflation (e.g.~\cite{Moroi:1994rs, Kawasaki:1995cy, Moroi:1999zb,Chen:2018uzu, Kitano:2008tk,Drees:2017iod}). Since the thermal bath is not significant during the scalar field dominated phase, the dominant sterile neutrino production happens during the radiation dominated phase, when $T<T_{RH}$~\cite{Gelmini:2004ah,Gelmini:2008fq}.

While our point is illustrated with the above scenarios, we note that many other cosmologies have been proposed, including kination~\cite{Spokoiny:1993kt,Joyce:1996cp,Salati:2002md,Profumo:2003hq,Pallis:2005hm} that may arise in models of quintessence, brane cosmology~\cite{Randall:1999vf} or ultra-fast expansion~\cite{DEramo:2017gpl}.  In future work~\cite{Gelmini:prep}, we will discuss sterile neutrino production within a phenomenological framework of cosmologies where entropy is conserved, which includes kination and ST as special cases.

In the DW~\cite{Dodelson:1993je} mechanism, sterile neutrino production is well described by a Boltzmann equation and is proportional to the ratio $\Gamma/ H$ of the conversion rate and the Hubble expansion rate.  As active neutrinos interact with the surrounding plasma they can convert to sterile neutrinos with some probability,  determined by the active-sterile mixing in matter due to the neutrino self energy. The production rate is usually not fast enough for sterile neutrinos to equilibrate and the process is a freeze-in of the final abundance. In the Std cosmology  the  momentum distribution $f_{\nu_s}$ of the sterile neutrinos is proportional to the momentum distribution of active neutrinos. In the LRT and ST model it has an extra $\epsilon$ and $\epsilon^{- 0.27}$ factor, respectively, where $\epsilon = p/T$ is the temperature-scaled dimensionless sterile neutrino momentum.

The maximum of the sterile neutrino production~\cite{Gelmini:prep}, specifically the maximum of $\left(\partial f_{\nu_s}(E,T)/\partial T\right)_\epsilon$, occurs in the standard cosmology at
\begin{equation}
\label{eq:Stdmax}
    T_{\textrm{max}}^{\rm Std} = 145 \textrm{ MeV} \left(\frac{m_s}{\textrm{keV}}\right)^{\frac{1}{3}}\epsilon^{-\frac{1}{3}}~.
\end{equation}
Since $T_{\textrm{max}}^{\rm Std}>T_{RH}=$5 MeV for all the sterile neutrinos we study, in the LRT cosmology  the production rate is never maximum, it is always smaller than in the Std. In the ST model the maximum of the production happens during the non-standard phase, at
\begin{equation}
\label{eq:st1tmax}
    T_{\textrm{max}}^{\rm ST} = 156 \textrm{ MeV} \left(\frac{m_s}{\textrm{keV}}\right)^{\frac{1}{3}}\epsilon^{-\frac{1}{3}}~.
\end{equation}

In the standard cosmology, the  ratio of the sterile  and active neutrino number densities $n_{\nu_s}/n_{\nu_{\alpha}}$, both computed at the same temperature $T$, is
\begin{align} 
\label{eq:numstd}
 \dfrac{n_{\nu_s}^{\rm Std}(T)}{n_{\nu_{\alpha}}(T)} \simeq 3.34 \times 10^{4}\sin^22\theta \left(\frac{d_\alpha}{1.13}\right)\left(\frac{30}{g_\ast}\right)^{\frac{3}{2}}\left(\frac{m_s}{\textrm{keV}}\right).
 \end{align}
  Here, $d_{\alpha} = 1.13$ for $\nu_{\alpha} = \nu_e$  and $d_\alpha = 0.79$ for $\nu_{\mu}$ and $\nu_{\tau}$,  $g_\ast$ is the entropy number of degrees of freedom at production, that we take to be $g_\ast= g_\ast (T_{\textrm{max}}^{\rm Std})$,   and $n_{\nu_{\alpha}}(T) = (3 \zeta(3)/2\pi^2)T^3 \simeq (1.804/\pi^2)T^3$.
  
  Assuming that only SM particles are present for $T$ higher than the QCD phase transition ($T \simeq 170$ MeV),  $g_\ast=80$ and approximately constant (it reaches 100 at 10$^5$ MeV). For $T$ values between close the QCD phase transition, the value of $g_\ast$ decreases steeply, and we take an average value of $g_{\ast} \simeq 30$ until $T$ decreases to $T= 20$ MeV. Between this temperature and $T =1$ MeV, when electrons and positrons become non-relativistic and annihilate, $g_{\ast} = 10.75$ (see e.g.~\cite{Husdal:2016haj,Borsanyi:2016ksw,Drees:2015exa}).

  In the ST cosmology the expansion rate $H$ is much larger, i.e. $\Gamma/ H$ is much smaller, than in the Std cosmology. Thus sterile neutrino production is suppressed, and the relic number density is much smaller than in the Std~\cite{Rehagen:2014vna, Gelmini:prep},
\begin{align}
 \label{eq:numst}
 \dfrac{n_{\nu_s}^{\rm ST}(T)}{n_{\nu_{\alpha}}(T)} \simeq&~ 0.54~ \sin^2 2\theta  \Big(\dfrac{d_{\alpha}}{1.13}\Big) \left(\frac{5\textrm{ MeV}}{T_{\rm tr}}\right)^{0.82}  \notag\\ &\times \left(\frac{30}{g_{\ast}}\right)^{\frac{3}{2}}  \left(\frac{m_s}{\textrm{keV}}\right)^{1.27}~,
\end{align}
where $g_\ast= g_\ast (T_{\textrm{max}}^{\rm ST})$. 

In the LRT cosmology, the dominant sterile neutrino production happens during the radiation dominated phase, at $T< T_{RH}$. Thus, if $T_{RH} < T_{\rm max}$, the  resulting relic sterile neutrino number density is smaller than in the Std~\cite{Gelmini:2004ah}
\begin{equation} 
\label{eq:numlrt}
\dfrac{n_{\nu_s}^{\rm LRT}(T)}{n_{\nu_{\alpha}}(T)} \simeq 11.3~\sin^2 2\theta \Big(\dfrac{d_{\alpha}}{1.13}\Big)  \Big(\dfrac{T_{RH}}{5~\text{MeV}}\Big)^3~.
\end{equation}

Notice that the temperature of sterile $T_{\nu_s}$ and active neutrinos $T_{\nu_\alpha}$ can be  different, due to entropy transfer into active neutrinos after the production of the bulk of the sterile neutrinos, in which  case  the right-hand side of Eqs.~\eqref{eq:numstd}  and~\eqref{eq:numst} needs to be multiplied by  $(T_{\nu_s}/ T_{\nu_\alpha})^3=10.75/ g_\ast$ to obtain the actual ratio of number densities. In the LRT model, most of the sterile neutrino production happens close to the reheating temperature, since the production rate deceases as $T$ decreases. Here $T_{\rm RH}$ is just 5 MeV, thus $g_\ast= 10.75$, and active neutrinos decouple at slightly lower temperatures of a few MeV when $g_\ast$ is the same, thus $T_{\nu_s}= T_{\nu_\alpha}$.

The  ratio of the energy density of relativistic neutrinos and their number density, $\rho_{\nu_s}(T)/ n_{\nu_s}(T)$, defines the sterile neutrino average momentum in each scenario $\langle p \rangle= \langle \epsilon \rangle T$ for each temperature~\cite{Gelmini:prep}, 
\begin{equation}
\label{eq:nonreseps}
\langle\epsilon\rangle = \dfrac{\rho_{\nu_s}(T)}{ T~ n_{\nu_s}(T)} 
= \left \{
  \begin{tabular}{cl}
  3.15, & ~~~\text{Std}\\
  2.89, & ~~~\text{ST} \\
 4.11, & ~~~\text{LRT}
  \end{tabular}
\right .\
\end{equation} 
which is used in the BBN and Ly-$\alpha$ limits below.

Comparing the number density ratios in
Eqs.~\eqref{eq:numstd}, \eqref{eq:numst} and   \eqref{eq:numlrt} it is clear  that  significantly larger mixing angles are allowed in the ST and LRT cosmologies than in the Std  by all upper limits dependent on the sterile neutrino relic abundance. In Fig.~\ref{fig:allDWlim} we display these constraints for the Std, ST and LRT cosmologies (in the left, middle and high panels, respectively),  extending the parameter space and updating the previous results of~\cite{Abazajian:2017tcc,Gelmini:2004ah,Rehagen:2014vna} (details of the calculations will be given in ~\cite{Gelmini:prep}). 

 All the sterile neutrinos we consider are non-relativistic at present, thus their  energy density is 
 $\rho_{s,0}= n_{\nu_{s,0}} m_s$ ($n_{\nu_{s,0}}$ is the present number density) which
must not be larger than the present DM energy density,  $\rho_{s,0} \leq \rho_{\rm DM}= \Omega_{\rm DM} \rho_c $, or
\begin{equation}
\Omega_s h^2 = \Big(\dfrac{m_s n_{\nu_{s,0}}}{\rho_c}\Big)h^2 \leq \Omega_{\rm DM} h^2 = 0.1186~.
\end{equation}
Here $\rho_c = 1.05 \times 10^{-5} h^2$ GeV/cm$^{3}$ is the critical density of the Universe and $h = 0.678$~\cite{Tanabashi:2018oca}. In Fig.~\ref{fig:allDWlim} we show the lines in mass-mixing space for several values of the $\Omega_s/ \Omega_{\rm DM}$ ratio, and indicate the forbidden region where this fraction is $>1$, assuming for simplicity  $g_\ast = 30$ for all the sterile neutrinos we consider.

If sterile neutrinos constitute the DM, their free-streaming would suppress structure formation below the free-streaming length  (see e.g.~\cite{Boyarsky:2018tvu})
\begin{equation}
\lambda_s \simeq 1~ \text{Mpc}\Big(\dfrac{\text{keV}}{m_s}\Big)
\Big(\dfrac{\langle \epsilon \rangle}{3.15}\Big) \Big(\dfrac{T_{\nu_s}}{ T_{\nu_\alpha}}\Big) ~,
\end{equation}
where $(T_{\nu_s}/ T_{\nu_\alpha})= (10.75/ g_\ast)^{1/3}$.  Observations  of  the  Lyman-$\alpha$ forest  in  the  spectra  of  distant  quasars  constrain  the  power spectrum on the $\sim$0.1-1 Mpc scales, thus imposing limits on neutrinos in the keV mass range, which would thus be WDM. These limits are usually given in terms of the mass  $m_{\rm th}$ of a fermion that was is thermal equilibrium and has a temperature $T_{\rm th}$ that depends on when it decoupled. Thus its mass and relic density (which depends on it temperature  and mass) are independent quantities.  Following~\cite{Viel:2005qj}, we equate the free-streaming scales  of this thermal WDM fermion and  a sterile neutrino (i.e. $\langle \epsilon \rangle T_{\nu_s}/m_s = 3.15~ T_{\rm th}/m_{\rm th}$) as well as their energy densities, to identify the  sterile neutrino mass $m_s$ that would result in the same free-streaming as a WDM fermion with mass $m_{\rm th}$ in thermal equilibrium. In this way we translate  the Lyman-$\alpha$ forest bounds~\cite{Baur:2017stq} from  $m_{\rm th}$ to $m_s$, which we display in Fig.~\ref{fig:allDWlim}. We extend the upper limit on the density fraction of the smallest tested masses,
as an upper limit on the density of lighter sterile neutrinos which would be Hot DM (the region rejected by this limit, shown in gray, is labeled  Ly-$\alpha$/HDM in Fig.~\ref{fig:allDWlim}),  until it is superseded by the BBN limit.
  
The impact on BBN of an increased expansion rate of the Universe  yields  an upper limit on  $N_{\rm eff}$, the effective number of relativistic active neutrino species present during BBN. Assuming that only sterile neutrinos and SM active neutrinos contribute  to $N_{\rm eff}$,
\begin{equation} 
N_{\rm eff} = 3.045+ (\rho_{\nu_s}/\rho_{\nu_\alpha})~,
\end{equation}
where $3.045$ is the contribution of the SM active neutrinos alone~\cite{Mangano:2005cc, deSalas:2016ztq}. All the sterile neutrinos we consider are relativistic during BBN, thus
 $(\rho_{\nu_s}/\rho_{\nu_\alpha})$ is the ratio during BBN of the relativistic energy densities of the sterile neutrino, $\rho_{\nu_s} = \langle\epsilon\rangle T n_{\nu_s}$ 
 with $ \langle\epsilon\rangle$ given in Eq.~\eqref{eq:nonreseps},  and one active neutrino species $\nu_{\alpha}$, $\rho_{\nu_\alpha} = 3.15 T n_{\nu_\alpha}$,   thus 
\begin{equation} 
\label{eq:neffeq}
\Delta N_{\rm eff} = N_{\rm eff} - 3.045 \simeq  \Big(\dfrac{\langle \epsilon \rangle}{3.15}\Big) \Big(\dfrac{n_{\nu_s}}{n_{\nu_{\alpha}}}\Big)\Big(\frac{10.75}{g_\ast}\Big)^{1/3}.
\end{equation}
Here the ratio ${n_{\nu_s}}/n_{\nu_{\alpha}}$ in each cosmology (see Eqs.~\eqref{eq:numstd}, \eqref{eq:numst} and   \eqref{eq:numlrt}) is the same during BBN as at present, since both  the number densities  just redshift for $T <$ 1 MeV (and BBN starts at about $T= 0.8$ MeV). 

 We use the BBN bound $N_{\rm eff} < 3.4$ at 95\% confidence level~\cite{Tanabashi:2018oca} in Eq.~\eqref{eq:neffeq} to set the limits we display in Fig.~\ref{fig:allDWlim} (the regions forbidden by this limit are shown in cyan).  This upper limit is similar in magnitude to the Planck 2018 limit~\cite{Aghanim:2018eyx} on $N_{\rm eff}$ derived from cosmic microwave background radiation (CMB) data. However, the CMB limit is applicable only to very light neutrinos ($m_s \ll 1$ eV) that are relativistic during recombination and hence we do not consider it. The  Planck 2018 and BICEP2/Keck and BAO data limits on the effective sterile mass $m_{s, {\rm eff}}$ or the sum of active neutrino masses~\cite{Aghanim:2018eyx, Choudhury:2018sbz} do not significantly change those we obtained based on $N_{\rm eff}$ during BBN.  Earlier Planck limits on $N_{\rm eff}$ and $m_{s, {\rm eff}}$ were considered in Ref.~\cite{Rehagen:2014vna}. 
 
 Photons produced in the decays of sterile neutrinos before recombination, i.e.~$\tau< t_{\rm rec}\simeq 1.2 \times 10^{13}$ sec, can  distort the CMB spectrum~\cite{Ellis:1990nb, Hu:1993gc} if they are produced after the thermalization time $t_{\rm th} \simeq 10^6$ sec, before which they could be incorporated into the  measured  Planck spectrum (see e.g. the discussion in Ref.~\cite{Gelmini:2008fq}). The COBE FIRAS limits~\cite{Fixsen:1996nj} on these distortions reject lifetimes $t_{\rm rec}> \tau > 10^6$ sec (region diagonally hatched in red in Fig.~\ref{fig:allDWlim} labeled CMB). Lifetimes of Majorana sterile neutrinos\footnote{For Dirac neutrinos the lifetime must be multiplied by 2.} equal to the lifetime of the Universe, $t_U= 4.36 \times 10^{17}$ sec, to $t_{\rm rec}$ and to $t_{\rm th}$ are indicated with red long-dashed straight lines in Fig.~\ref{fig:allDWlim}.  
 
 The energy loss due to sterile neutrinos produced in core collapse supernovae explosions disfavors~\cite{Kainulainen:1990bn} the region
 horizontally hatched in brown and labeled SN  in Fig.~\ref{fig:allDWlim}. Due to the considerable uncertainty in the neutrino transport and flavor transformation within hot and dense nuclear matter~\cite{Abazajian:2001nj}, it is difficult to exclude this region entirely. For recent studies regarding sterile neutrinos mixing with
 $\nu_\mu$ or $\nu_\tau$ see e.g.~Ref.~\cite{Raffelt:2011nc,Arguelles:2016uwb}.

The most restrictive limits on sterile neutrinos with $m_s\,>\,1$ keV  come from astrophysical indirect detection searches of X-rays produced in the $\nu_s \rightarrow \nu_\alpha \gamma$  two-body decay~\cite{Ng:2019gch,Perez:2016tcq,Neronov:2016wdd}. The rate of this decay mode is~\cite{Shrock:1974nd, Pal:1981rm}
\begin{equation}
\label{eq:xraydecay}
    \Gamma_\gamma = 1.38\times10^{-32} ~\text{s}^{-1} \left(\frac{\sin^2 2\theta}{10^{-10}}\right)\left(\frac{m_s}{\textrm{keV}}\right)^{5}~.
\end{equation}
Due to the rapid decrease of this rate with decreasing mass, X-ray observations do not provide meaningful constraints on sterile neutrinos with $m_s< 1$ keV. The  model independent X-ray bounds derived from observations of galaxies and clusters~\cite{Ng:2019gch,Perez:2016tcq,Bulbul:2014sua,Boyarsky:2014jta}   assume that sterile neutrinos constitute the entirety of the DM. To translate  these limits into those that apply in our scenarios we note
that the X-ray signal depends on the produced photon flux, which is proportional to
the product $ (\Omega_s/ \Omega_{\rm DM}) \sin^22\theta$.
The present sterile neutrino fraction  $(\Omega_s/ \Omega_{\rm DM})$ of the DM is itself proportional to $\sin^2 2\theta$ (due to the dependence of ${n_{\nu_s}}$ in Eqs.~\eqref{eq:numstd}, \eqref{eq:numst} and   \eqref{eq:numlrt} on the mixing).  Hence, the X-ray limits shown in green is Fig.~\ref{fig:allDWlim} (labeled Xray) are related to the published model independent bounds  through a simple  rescaling.   The same rescaling applies to diffuse extragalactic background radiation (DEBRA) bounds~\cite{Boyarsky:2005us} on sterile neutrinos decaying after recombination, i.e. $\tau> t_{\rm rec}$ (the lower boundary of the DEBRA rejected region is indicated by a straight green line in Fig.~\ref{fig:allDWlim}). 

The $3.5$ keV X-ray emission line signal reported in 2014~\cite{Bulbul:2014sua,Boyarsky:2014jta}, which remains a matter of lively debate~(see e.g.~\cite{Dessert:2018qih,Boyarsky:2018ktr}),  could be due to the decay of $m_s\simeq$ 7 keV sterile neutrinos whose mixing should be $\sin^22\theta=5\times10^{-11}$ if they constitute all of DM.
The active-sterile mixing  necessary  for sterile neutrinos  produced non-resonantly to reproduce the putative signal line shifts in the same way as the X-ray bounds just mentioned, and  is indicated with a black star in Fig.~\ref{fig:allDWlim}. 
Notice in Fig.~\ref{fig:allDWlim} that the signal would correspond to sterile neutrinos that constitute only a small fraction of the DM. It is clear that in  pre-BBN cosmologies in which the sterile neutrino relic number density is smaller the  mixing angle required to produce the signal is larger, which potentially brings it within easier reach of laboratory experiments. In particular,  in the LRT and ST cosmologies shown  Fig.~\ref{fig:allDWlim} the necessary mixings are  $\sin^2 2 \theta = \mathcal{O}(10^{-7})$, within  reach of the KATRIN experiment with its proposed TRISTAN upgrade~\cite{Mertens:2018vuu} as well as   the upcoming HUNTER experiment and its upgrades~\cite{Smith:2016vku}.

Since laboratory experiments directly probe the active-sterile mixing angle and mass by searching for active neutrino appearance and disappearance, the resulting bounds they set are independent of cosmology. The tritium decay KATRIN experiment~\cite{Wolf:2008hf,Mertens:2015ila} has started data collection. The sterile neutrino mass eigenstate would manifest itself as a kink-like distortion of the measured $\beta$-decay spectrum.  In the keV-scale mass range it will probe mixings down to $\sin^2\theta \sim 10^{-4}$~\cite{Mertens:2018vuu}. The upgrade
 of the experiment, TRISTAN, is expected to reach sensitivities of  $\sin^2 \theta \sim  10^{-8}$ 
 in 3 years of run time~\cite{Mertens:2018vuu}. An even more impressive sensitivity is expected to be reached in the keV mass-range   by the upcoming cesium trap experiment HUNTER \cite{Smith:2016vku}. In this experiment the neutrino missing mass will be reconstructed from $^{131}$Cs electron capture  decays occurring in a magneto-optically trapped sample. The initial prototype, currently under construction, will have two phases A and B, with a final reach denoted by H1 in Fig.~\ref{fig:allDWlim}, whose goal is to demonstrate the feasibility of the experimental technique. In the HUNTER upgrade, denoted by HU in Fig.~\ref{fig:allDWlim}, the sensitivity is projected to reach 
 $\sin^2 2 \theta \sim 10^{-11}$. 
   
 For eV-scale mass sterile neutrinos, a slew of reactor and accelerator neutrino experiments probe and restrict portions of the parameter space. The combined reactor neutrino constraints from the Daya Bay~\cite{An:2016luf}, Bugey-3~\cite{Declais:1994su} and  PROSPECT~\cite{Ashenfelter:2018iov} experiments are displayed in Fig.~\ref{fig:allDWlim} (green region labeled R). Several anomalous results, consistent with the existence of a eV-scale mass sterile neutrino, have been reported. In particular, long-standing excesses observed within $\nu_e$ appearance channels  in the short-baseline LSND~\cite{Aguilar:2001ty} and MiniBooNE~\cite{Aguilar-Arevalo:2013pmq} experiments have been recently strengthened with additional MiniBooNE data~\cite{Aguilar-Arevalo:2018gpe}. We display in Fig.~\ref{fig:allDWlim}  the black hatched band denoted MB in sterile neutrino parameter space consistent at the 4-$\sigma$ level with these excesses taken from Fig.~4 of~\cite{Aguilar-Arevalo:2018gpe} (see~\cite{Aguilar-Arevalo:2018gpe} for details).  We note, however, that the sterile neutrino interpretation of these appearance excesses is in strong tension with $\nu_{\mu}$ disappearance data from IceCube~\cite{TheIceCube:2016oqi} and MINOS+~\cite{Adamson:2017uda}. 
  
  Additionally, combined fits~\cite{Dentler:2018sju,Gariazzo:2018mwd,Liao:2018mbg} to recent reactor neutrino results by the DANSS~\cite{Alekseev:2018efk} and NEOS~\cite{Ko:2016owz} experiments are consistent with an interpretation based on  a sterile neutrino with $m_s \simeq 1.14$ eV mass and  $\sin^2 2\theta \simeq 0.04$ mixing with the $\nu_e$ active neutrino. The regions allowed at 3-$\sigma$ by DANSS and NEOS data  are shown with black vertical elliptical contours (reproduced from Fig. 4 of~\cite{Gariazzo:2018mwd}). Further, as we also show in Fig.~\ref{fig:allDWlim},  the PTOLEMY~\cite{Betti:2019ouf} and KATRIN~\cite{megas:thesis}  experiments will be able to test all or part of this parameter space. Note
  in Fig.~\ref{fig:allDWlim}, that 
  in the ST and LRT cosmologies the cosmological bounds on eV-scale mass sterile neutrinos  are significantly relaxed with respect to those in the Std cosmology, and the parameter space required for the above excesses is not rejected by them. 
  
  If neutrinos are Majorana fermions, neutrinoless double-beta ($0\nu\beta\beta$) decays are allowed and their half-life times depend on an effective electron neutrino  Majorana mass $\left<m\right>$. 
  A sterile neutrino mixed with  $\nu_e$
 contributes  a component $\left<m\right>_s = m_s e^{i\beta_s} \sin^2 \theta$ to this effective mass, where $\beta_s$ is a Majorana CP-violating phase. Using the present bound on the magnitude of the effective Majorana mass, $|\langle m\rangle|<0.165$ eV~\cite{KamLAND-Zen:2016pfg}, we obtain $m_s\sin^2 2\theta<0.660$ eV.
This bound labeled $0\nu\beta\beta$ is shown by a straight orange line in Fig.~\ref{fig:allDWlim}.

In summary, we revisited non-resonant sterile neutrino production in different pre-BBN cosmologies, before the temperature of the Universe was 5 MeV and showed that these neutrinos can act as sensitive probes of these cosmologies. This highlights that cosmological bounds are not robust and direct laboratory searches, whose results are independent of cosmology, are particularly important to test both, particle physics and cosmology.  The pre-BBN epoch of the Universe has not yet been tested because of the lack of any detected remnant from it.  A visible sterile neutrino could be the first.

In particular in the LRT and ST models  we studied here,
cosmological bounds are significantly relaxed, and the mixing of the sterile neutrinos possibly responsible for the 3.5 keV signal are within  easier
reach of the upcoming KATRIN/TRISTAN and  HUNTER experiments. These experiments have already started or are expected to start taking data soon. The observation of a 7 keV mass sterile neutrino in one of them would not only constitute the discovery of a new elementary particle,  a particle physics discovery of fundamental importance, but could hold vital information about the pre-BBN cosmology from which this sterile neutrino would be the first ever detected remnant. For example, if the measured mixing is found to be $\sin^2 2 \theta = \mathcal{O}(10^{-7})$, the discovery would be consistent with a  non-standard cosmology such as the ST and LRT we studied here (or maybe special particle models, e.g.~\cite{Bezrukov:2017ike}) and if $\sin^2 2 \theta = \mathcal{O}(10^{-10})$ it would point to a  standard pre-BBN cosmology, as indicated in Fig.~\ref{fig:allDWlim} (see the location of the black stars).

Furthermore, the relaxation of cosmological bounds on eV-mass scale sterile neutrinos in non-standard pre-BBN cosmologies allows for the results reported from the LSND, and MiniBooNE short-baseline as well as the DANSS and NEOS reactor neutrino experiments to be unrestricted by cosmology. If the discovery of a sterile neutrino in any of these experiments would be confirmed, again this would be of fundamental importance not only for particle physics, but for the pre-BBN cosmology in which they were produced, and it could provide an indication of a non-standard cosmology in this yet untested epoch.
  
We would like to thank Eric Hudson, Bedrich Roskovec, Michael Smy and Hank Sobel for comments.
This work was supported in part  by the U.S. Department of Energy (DOE) Grant No. DE-SC0009937.
 
\bibliography{sternucos}

\end{document}